# LURAD: Design Study of a Comprehensive Radiation Monitor Package for the Gateway and the Lunar Surface

C. Potiriadis[2,1] (constantinos.potiriadis@eeae.gr), K. Karafasoulis[3,1] (ckaraf@gmail.com), C. Papadimitropoulos[1] (christos.papadimitropoulos@gmail.com), E. Papadomanolaki[4] (e.papadomanolaki@adveos.com), A. Papangelis[1] (apapange@aerospace.uoa.gr), I. Kazas[6,1] (kazas@inp.demokritos.gr), J. Vourvoulakis[5,1] (jvourv@ihu.gr), G. Theodoratos[4] (g.theodoratos@adveos.com), A. Kok[8] (Angela.kok@sintef.no), L. T. Tran[7] (thuy_linh_tran@uow.edu.au), M. Povoli[8] (marco.povoli@sintef.no), J.Vohradsky[7] ( jamesv@uow.edu.au), G. Dimitropoulos[4] (g.dimitropoulos@adveos.com), A. Rosenfeld[7] (anatoly@uow.edu.au), C. P. Lambropoulos[1+] (lambrop@uoa.gr)

+Corresponding author

[1]Department of Aerospace Science and Technology, National and Kapodistrian University of Athens, Thesi Skliro, Psahna-Evia, 34400 Greece

[2]Greek Atomic Energy Commission, Agia Paraskevi, Attiki, 15310 Greece

[3]Hellenic Army Academy, Evelpidon Avenue, Vari, 16673 Greece

[4]ADVEOS Microelectronic Systems P.C., 256 Mesogeion Avenue, Athens, Attiki, 15451 Greece

[5] Hellenic International University, Serres, 62124 Greece

[6]Institute of Nuclear and Particle Physics (INPP), NCSR Demokritos, Agia Paraskevi, Attiki, 15310 Greece

[7]Centre for Medical Radiation Physics, University of Wollongong, NSW, Australia

[8]SINTEF, SINTEF MiNaLab, Oslo, Norway

**Abstract**

Moon is an auspicious environment for the study of Galactic cosmic rays (GCR) and Solar Particle Events (SEP) due to the absence of magnetic field and atmosphere. The same characteristics raise the radiation risk for human presence in orbit around it or at the lunar surface. The secondary (albedo) radiation resulting from the interaction of the primary radiation with the lunar soil adds an extra risk factor, because neutrons are produced, but also it can be exploited to study the soil composition. In this paper, the design of a comprehensive radiation monitor package tailored to the lunar environment is presented. The detector, named LURAD, will perform spectroscopic measurements of protons, electrons, heavy ions, as well as gamma-rays, and neutrons. A microdosimetry monitor subsystem is foreseen which can provide measurements of LET(Si) spectra in a wide dynamic range of LET(Si) and flux for SPE and GCR, detection of neutrons and biological dose for radiation protection of astronauts. The LURAD design leverages on the following key enabling technologies: (a) Fully depleted Si monolithic active pixel sensors; (b) Scintillators read by silicon photomultipliers (SiPM); (c) Silicon on Insulator (SOI) microdosimetry sensors; These technologies promise miniaturization and mass reduction with state-of-the-art performance. The instrument's design is presented, and the Monte Carlo study of the feasibility of particle identification and kinetic energy determination is discussed.

# 1    INTRODUCTION

The need for a radiation detector which could monitor and characterize the GCR and SEP particles and the secondary radiation from the lunar surface has been emphasized by international science teams in the payload investigations for the Lunar Orbital Platform – Gateway (Dandouras et al., 2023; Losekamm and Berger, 2018). The first science payloads to be flown on the Gateway are the European Radiation Sensors Array (ERSA) and the Heliophysics Environmental and Radiation Monitoring Experiment Suite (HERMES) instruments suites both consisting of collections of devices with high technology readiness level (TRL) and flight heritage. These payloads have particles and fields instrumentation for radiation and space weather studies. The European Space Agency has initiated the discussion for the successor ERSA-2.

The kinetic energy spectrum of GCR protons extends from a few MeV up to $10^{20}$ eV. The low energy part of this spectrum is sensitive to the magnetic field of the heliosphere and has not been directly measured yet, because the existing GCR observatories are in low earth orbit, where this part is filtered out. The measurement of fast neutron and gamma albedo particle fluxes with the aid of an instrument in low lunar orbit or on the lunar surface is expected to reveal variations in the elemental composition of the lunar regolith (Zaman et al., 2021).

Besides, space radiation is one of the main health risks for human exploration of the Solar system. NASA categorizes it as the first and most menacing out of five hazards for a travel to MARS (https://www.nasa.gov/hrp/5-hazards-of-human-spaceflight). The measurements of the Mars Science Laboratory during travel to Mars (Zeitlin, 2013) and on the surface of the planet (Hassler et al., 2014) have proven that the dose for typical mission scenarios approaches 1 Sv, while the LND measurement on the lunar surface is 1.37 ± 0.25 mSv·day-1, about 2.6 times higher than the International Space Station crew's daily dose (Zhang et al, 2020). It is also well known that the understanding of the risk posed to long-duration astronauts by the radiation environment in space is limited, despite years of research (Chancellor et al., 2018). The uncertainty on the radiation risk is still very high, and countermeasures are not readily available. Consequently, the complete description of the radiation field which is provided by the fluence spectrum as a function of charge and energy of the impinging particles is of high importance because this is the input to all the risk assessment algorithms either existing or new (NCRP, 2007).

The LURAD instrument is designed with purpose to identify protons, ions, neutrons, electrons/positrons, and gamma rays, and measure their kinetic energy. In addition, it will provide dosimetry information and alerts for potentially hazardous events. The simulated performance of the instrument is summarized in Table 1. No magnet is used, resulting in a low mass footprint. Nevertheless, particle identification and kinetic energy determination well above 500 MeV/n is feasible.

LURAD leverages on fully depleted Si monolithic active pixel sensors (Lambropoulos C.P. et al., 2019), and scintillators read by silicon photomultipliers (SiPM). The microdosimetry module is based on the Silicon on Insulator sensors (Kok A et al. 2020, Pan V. et al. 2023, S. Perrachi et al. 2020). The techniques used are energy deposition measurements on plastic and crystalline scintillators, and on Si active pixel sensors; Time-of-flight (ToF) measurements with the aid of fast scintillators are used to extend the kinetic energy measurement capability towards the GeV range. Multivariate analysis (MVA) methods have been applied to investigate LURAD performance into the radiation fields where it is supposed to measure and operate.

Section 2 presents a concise description of the instrument. The analysis that led to the results summarized in Table 1 is presented in Sections 3 and 4. Section 5 discusses some limitations of the study related to the present status of LURAD development and presents the main conclusions from the reported work.

Table 1: Simulated measurement capabilities

| |
|---|
| Protons and ions spectra in the energy range from 40 MeV/u up to 2 GeV/u |
| Discriminate GCR particle species with atomic number from Z=1 up to Z=26 |
| Electrons spectrum in the energy range from < 1MeV up to 20 MeV |
| gamma photons spectrum in the energy range from 100 keV up to 10 MeV |
| Neutrons spectra in the energy range from 100 keV to 300 MeV. |
| Microdosimetry measurements and early warning capability |

## 2 DESCRIPTION OF LURAD

LURAD incorporates three measuring sub-systems: (a) Proton-Ion-Fast Neutron with the task to identify protons, ions, and fast neutrons and measure their kinetic energy. (b) The Electron-Gamma subsystem with the task to identify electrons and gamma photons. (c) The Microdosimetry subsystem with the task to provide time-resolved measurements of Total Ionizing and Non-Ionizing Dose (TID, TNID), micro-dosimetry and LET spectra, and biological dose. The measuring subsystems send data to and receive commands by a central processor board. This board, embedded in the LURAD package, transmits the science data to an external transceiver. A power supply board provides regulated voltages to all subsystems. The subsystems are described in the following paragraphs. The instrument's mass has been calculated to be ≈3 kg using the detailed dimensions and materials. Its shape and external dimensions are shown in Figure *1*. The current estimation for the power consumption is around 15 W. Breadboards for the different subsystems have been developed.

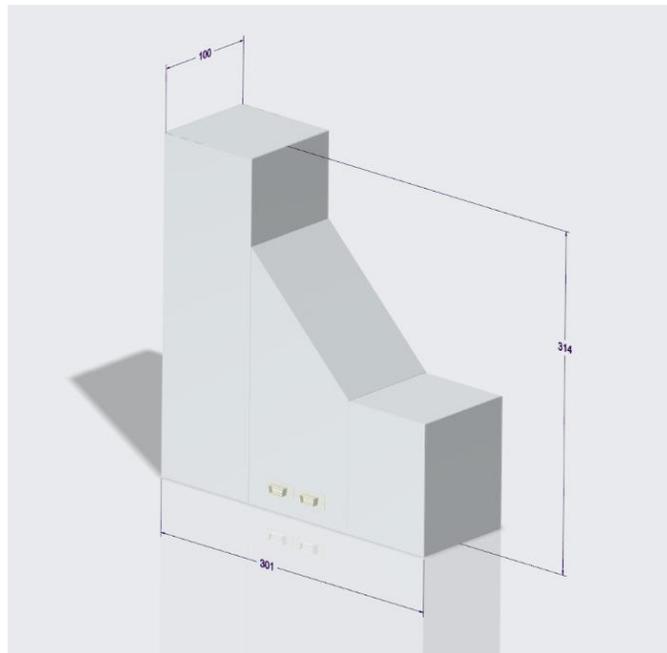

Figure 1: LURAD shape and external dimensions in mm.

## 2.1 Proton-Ion-Fast Neutron sub-system

In the center of the proton-ion-fast neutron subsystem a plastic cubic scintillator enables the transfer of the neutron kinetic energy to its hydrogen nuclei. This cubic scintillator, as a low Z material has low efficiency for interaction with the gamma photons. The central cube is surrounded by sensitive volumes which act as vetoing detectors in case of neutral particles or as active detectors in case of charged particles. Figure *2* illustrates the design of the subsystem. It is a telescope of 30 cm length with a cross section area of 5x5 cm$^2$. The telescope is symmetric along its axis and consequently it can measure particles entering either from the top or the bottom face. The top and bottom faces are covered by 0.1 mm thick aluminum. Behind the covers are placed two EJ232Q fast scintillators with dimensions 5x5x0.5 cm$^3$ each (labelled A in Figure *2*). Next to the EJ232Q, a layer of Silicon active pixel sensors array is placed (labelled B2 in Figure *2*). The Silicon active pixel sensors array consists of 2x2 groups of 2x2 chips, each one with 100x100 pixels. Thus, a total area of 4x4 cm$^2$ is covered by 160000 pixels. These pixel sensors are of the "high gain" flavor and are described in section 2.4.

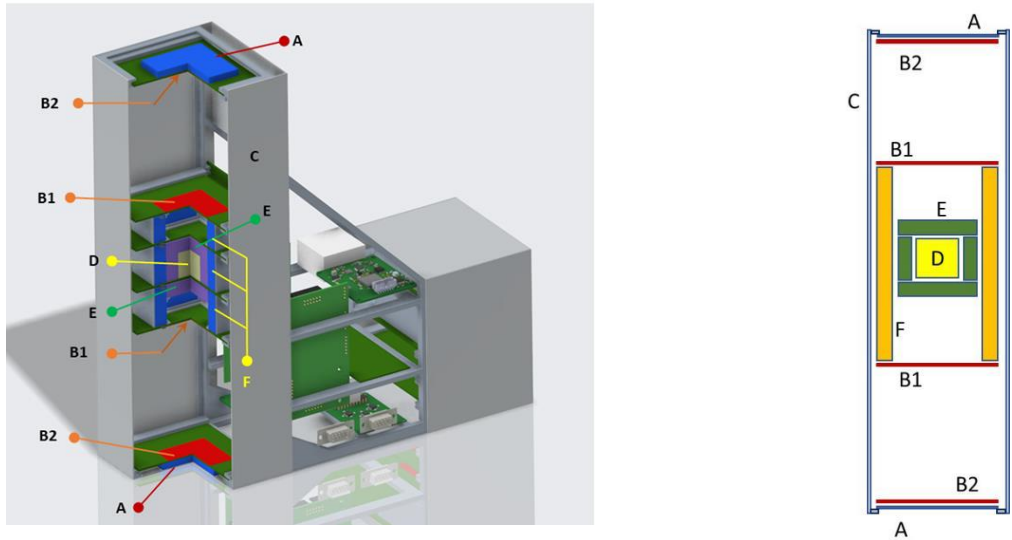

Figure 2: Γ cut (left) and simplified cross section (right) of the proton-ion-fast neutron sub-system

The center of the proton-ion-fast neutron subsystem is occupied by a cube with dimensions 4x4x4 cm$^3$. The cube has an EJ200 plastic scintillator core with dimensions 2x2x2 cm$^3$ (labelled D in Figure *2*). The core is surrounded by six CsI(Na) scintillators 1cm thick (labelled E in *Figure 2*) . The lateral surfaces of the cube are covered by another layer of 1 cm thick EJ200 scintillators which extend beyond the top and bottom faces of the core by 2.5 cm. As can be seen in *Figure 2*, they are labeled F. They are twelve EJ200 scintillators organized in six blocks for the readout of their signal, and they constitute the anticoincidence detector of the telescope. Two layers of Silicon active pixel sensors arrays are placed on the top of the lateral EJ200 scintillators (labelled B1 in *Figure 2*). The B1 Silicon active pixel sensors are organized again in 2x2 blocks of 2x2 chips, but with 50x50 pixels with 200 µm pitch. Consequently, an area of 4x4 cm$^2$ is covered by 40000 pixels. The B1 sensors are of the "low gain" flavor (see Section 2.4 ). The lateral surfaces of the telescope enclosure are made of 1 mm thick aluminum. The EJ232 scintillators comprise the time-of-flight detectors of the subsystem. The pixel detector layers B1 and B2 will provide the necessary information to reconstruct the particle's track whose time of flight will be measured. Some details of the time-of-flight readout are provided in Section 2.5. The proton-ion-fast neutron subsystem incorporates SiPMs for the CsI(Na) crystals and the EJ200 plastic scintillators. Their signal outputs are collected in 13 readout channels. The signals are processed and digitized with the aid of the IDE3380 multichannel application specific integrated circuit (ASIC) by IDEAS. The IDE3380 ASIC sends address and digitized light output data to the field programmable gate array (FPGA) which controls the subsystem.

## 2.2 ELECTRON GAMMA SUBSYSTEM

In the center of the electron gamma subsystem, a CsI cubic crystal is placed, which, as a high Z material, is an effective gamma-ray detector. The three successive layers of pixelated Si detectors are used for electron identification. Figure *3* illustrates the subsystem. It is designed to measure particles coming from the one side (Figure *3*: side "towards the lunar surface"). It consists of a CsI(Na) scintillator with dimensions 4x4x4 cm$^3$ (Figure *3*: ScID=1) surrounded by five anticoincidence CsI(Na) detectors (Figure *3*: ScID=2, 3, 4, 5). The PixID=3 Si pixel sensors layer is placed on the face of the CsI(Na) scintillator and is of the "high gain" flavor. The PixID=2 is a low gain Si pixel sensors layer placed at 2 cm from the PixID=3 along the same direction. The PixID=1 Si pixel sensors layer is again of the "high gain" flavor, and it is placed at 2 cm from PixID=2 along the same direction (towards the lunar surface in Figure *3*). The active surface area of each layer is almost 4 x 4 cm$^2$ organized in the same way as the proton-ion-fast neutron sub-system. The ScID=2 CsI(Na) scintillator is covered with a Tantalum sheet 1 mm thick. The aluminum enclosure of the sub-system is 1 mm thick except for the face in front of the PixID=1 layer which is 0.1mm thick.

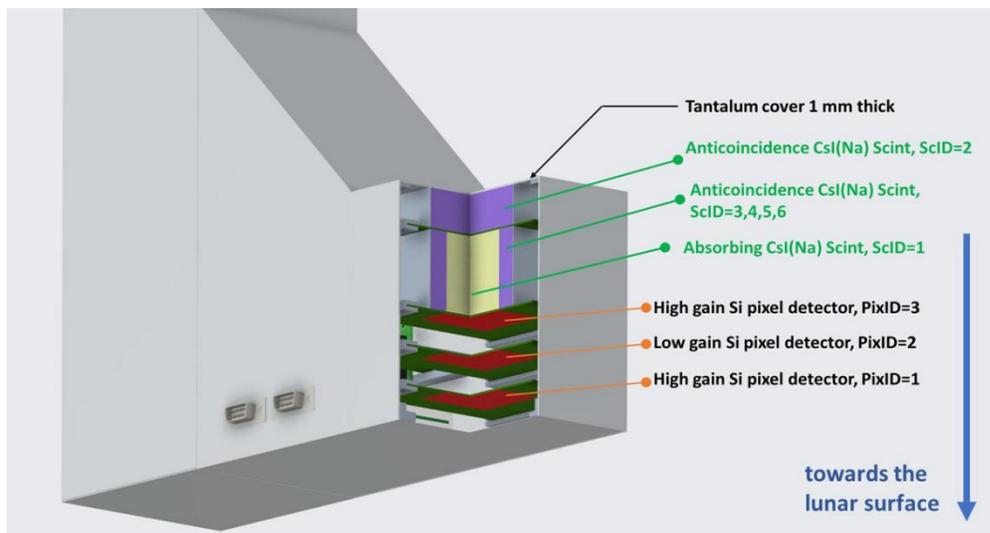

Figure 3: Γ cut of the electron – gamma sub-system. When the package is on the lunar surface (or in low lunar orbit) its bottom side should face the lunar surface.

In this subsystem also an IDE3380 ASIC receives the signals of the SiPMs attached to the corresponding CsI(Na) crystal blocks in six readout channels and sends address and digitized light output data to the FPGA which controls the subsystem.

## 2.3 MICRODOSIMETRY MODULE

The dose equivalent H is the main operational quantity used to assess the effective dose. It is obtained by multiplying the dose by the quality factor Q, of the radiation. Identification and kinetic energy determination of the GCR and SEP particles entering LURAD can be used to infer the quality factor. Microdosimetry is an alternative, useful approach for evaluating the quality factor in a mixed radiation field found typically in space, without knowing the energy or type of particles. In microdosimetry the stochastic energy depositions in sensitive volumes (SVs) of micrometric scale mimicking the cell are measured. Each SV is a single p-n junction diode with cylindrical shape with a n+ planar electrode surrounded by a p+ trench electrode. It is fabricated on a p-type high resistivity substrate. The cylindrical SV stays on top of a 100-µm-thick supporting substrate, isolated by a thin layer of silicon dioxide (Kok A et al. 2020).

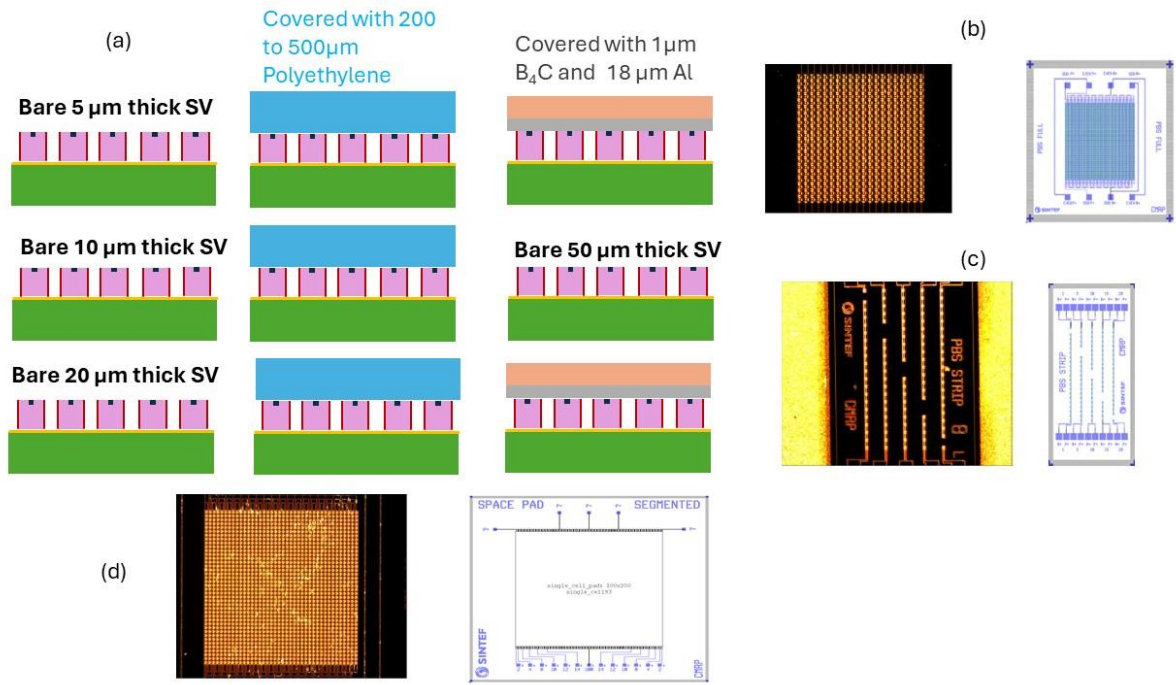

Figure 4: (a) The 9 PBS-FULL SOI sensors organized in 3 blocks. (b) Microphotograph and layout of the PBS-FULL SOI sensor. (c) Microphotograph and layout of the PBS-Strip SOI sensor. (d) Microphotograph and layout of the SPACEPAD SOI sensor.

The microdosimetry module encapsulates 11 SOI sensors. Nine of them named PBS-FULL (Pan V.A. et al., 2023) have 40 x 40 SV each, divided in two blocks of 800 SV. The SV of each block are read in parallel resulting to 18 readout channels for the nine sensors. The PBS-FULL sensors are divided in 3 groups of 3 sensors each (Figure 4a). One group is covered with $B_4C$ to become sensitive to thermal neutrons. A second group is covered with polyethylene to become sensitive to fast neutrons, while the third group is bare, and measures charged particles. The sensors, members of each group, have SV with height 5 μm or 10 μm, 20 μm and 50 μm and in this way they cover the linear energy transfer (LET) regions from 2 keV/μm to 58 MeV/μm (5 μm height), 1 keV/μm to 29 MeV/μm (10 μm height), 0,5 keV/μm to 14.5 MeV/μm (20 μm height), 0.2 keV/μm to 5.8 MeV/μm (50 μm height). One sensor named PBS_Strip (*Figure 4*c) is tailored to measure high flux SEP events, because it is divided to 5 readout channels for 1, 5, 10, 15 and 20 SV respectively. The last sensor is named SPACEPAD (*Figure 4*d) and consists of 200 x 200 SV with segments of 100, 200, 400, 500, 600, 700 and 5000 SV read separately in 7 channels. SPACEPAD is intended to measure the GCR field. Consequently, a readout application specific integrated circuit (ASIC) with 32 channels is used. Each sensor type has different number of SVs to cover different ranges of proton flux. The total proton flux which can be measured is up to $1.8·10^9$ $cm^{-2}·s^{-1}$. The worst 5 min average flux of protons with energy > 10 MeV for the 19 Oct 1989 SEP event calculated using CRÈME-96 is $4,35·10^{11}$ $m^{-2}·sr^{-1}·s^{-1}$. It would induce a mean count rate of 800 Hz in a single SV. This count rate can be easily handled by the electronic readout channels connected to the PBS_Strip sensor. On the other hand, SPACEPAD will provide about 480 counts in one min for a flux of 10 $protons/cm^2·s^{-1}·sr^{-1}$ (63 $protons/cm^{-2}·s^{-1}$) with energy > 10 MeV (LET< 8keV/μm(Si)). This will enable the system to provide early warning for a possible SEP event.

## 2.4 ACTIVE SILICON PIXEL SENSORS

Following the initial study (Peric, 2007), fully depleted Si monolithic active pixel sensors (called HVCMOS or DMAPS) have been introduced in high energy physics experiments for measuring the tracks of minimum ionizing particles. Fully depleted Si pixels enable fast and less prone to radiation damage charge collection. Today, research on achieving full integration of sensing and microelectronics in monolithic CMOS pixel sensors has been identified by the European Committee for Future Accelerators (ECFA) as one of the themes of its global Detector R&D Roadmap (ECFA, 2021). Coarsely segmented Si diode detectors are used in almost all instruments measuring the mixed radiation fields of the space environment. With starting event the installation of 5 Timepix units onboard the International Space Station in 2012 (Pinsky L., et al.,2014), hybrid Si pixel detectors have started to penetrate in space dosimetry. HardPix, a device using Timepix3 bump bonded to Si pixel sensors is included in the ERSA package. The introduction of fully depleted Si monolithic active pixel sensors in space instrumentation could enable high field of view with low mass footprint, deliberate granularity selection, and very low cost. The problems to be solved for measuring GCR and SEP events are the dynamic range of the charge signal which exceeds 80 dB, the event rates starting from 2 to 4 per $cm^2 \cdot s^{-1}$ and reaching $55 \cdot 10^7$ per $cm^2 \cdot s^{-1}$ (for a 4π sr field of view), and, most importantly, the fact that limited power is available from solar or other sources while the heat generated requires material (i.e., extra weight) to manage. With these considerations, we have selected a semiconductor fabrication technology which has been successfully used for high energy physics detectors (Barbero,2020). We developed two kinds of pixel architecture: The low gain pixel can process high energy (charge) depositions up to 8 pC in a 200 x 200 x 50 μm³ volume and the high gain can process low energy depositions by fast charged particles down to 0.5 fC in a 100 x 100 x 100 μm³. The prototype sensors have a lower number of pixels than foreseen. The pixel design specifications are collected in Table 2.

Table 2: Fully depleted monolithic active pixel sensor specifications

|  | Low gain pixel | High gain pixel |
|---|---|---|
| pixel size | 100 x 100 μm² | 200 x 200 μm² |
| Charge (Qin) range | 40 fC – 9 pC | 0.5 fC – 50 fC |
| Gain | 109 mV/pC | 17.5 mV/fC (Qin>3fC) 120 mV/fC (Qin<3fC) |
| Idle pixel current | 35 nA@1.8V | 7.5 μA@1.8V |
| Equivalent Noise Charge (ENC) | 1.5 fC | 200 aC |

In both sensor array versions, the voltage proportional to the charge collected is transferred to the array bottom and digitized. The digitized data are sent outside the chip using the Serial Peripheral Interface (SPI) protocol. The wired OR of the hit flags informs the external controller that a chip has received signal above threshold in one or more pixels. The chip readout is performed sequentially for the pixels which have raised their hit flags.

The low gain sensor has been fabricated and currently is experimentally evaluated. The high gain sensor has been submitted to the semiconductor foundry for fabrication.

## 2.5 THE TIME-OF-FLIGHT MEASUREMENT SYSTEM

A block diagram of the time-of-flight measuring system is presented in Figure *5*. The EJ232 fast scintillators are read by AFBR-S4N33C013 SiPMs and their outputs are amplified by custom discrete amplifiers. The amplified signals are driven to custom constant fraction discriminators. The discriminators provide trigger signals to a time to digital converter (TDC) which measures the time interval between them. The TDC is realized in an FPGA using the multiple tapped delay line architecture (Wu and Shi, 2008).

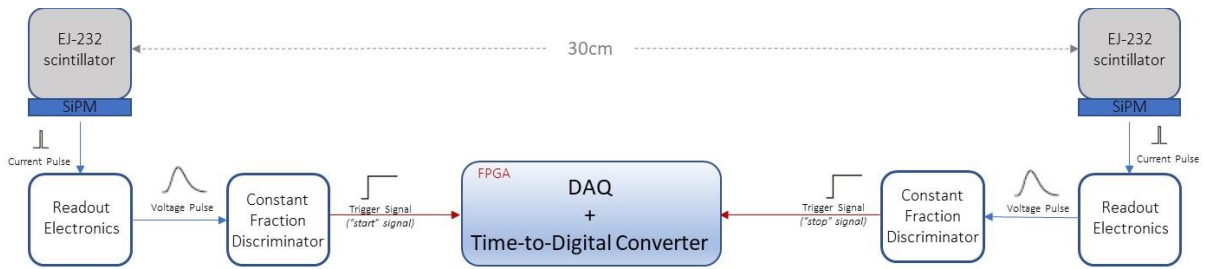

Figure 5: Block diagram of the time-of-flight system of LURAD

The development targets $\sigma = 30\ ps$ or even better time measurement resolution. The reported state of the art coincidence time resolution using a similar setup is of the order of 10 ps (Stoykov A. and Rostomyan T., 2021). The relation of the time measurement resolution to the kinetic energy estimation is discussed in paragraph 4.1.3. As can be seen in Figure 10, with 30 ps resolution proton kinetic energy spectra well above 1 GeV can be reconstructed.

## 2.6 CENTRAL PROCESSOR AND POWER SUPPLY

Each subsystem has a controlling FPGA which communicates with the central controller using the SPI based SPI4SPACE type 2 protocol. The central microprocessor is the SPI host device while the subsystem FPGAs constitute the SPI client devices. The central microprocessor selected is the space qualified GR716 by Gobham Gaisler. The central microprocessor is responsible for transmitting to an external transponder raw data, alert signals, or any of the data products of the device, implementing a SpaceWire node. A 50W isolated flyback DC/DC converter was designed to provide power to the instrument sub-systems. The power supply receives regulated input 28V and produces 5V/10A output. Several voltage levels are required on each board of the instruments. Most of them are created on each board. Figure *6* shows the placement of these units and of the microdosimetry subsystem in LURAD.

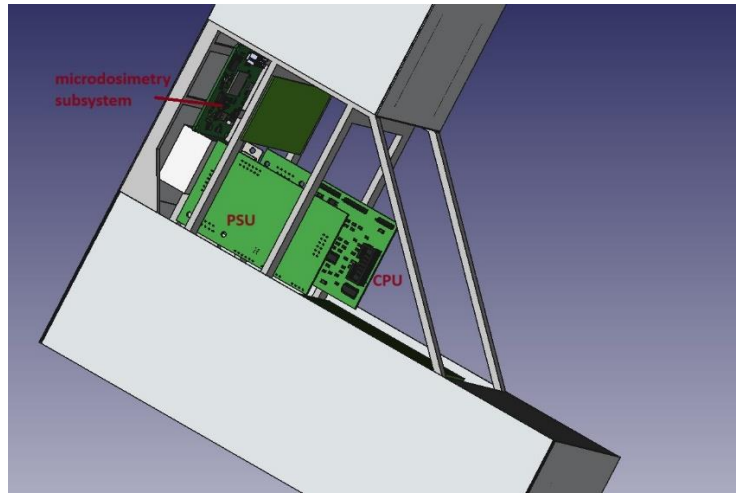

Figure 6: Placement of the central processor, of the power supply unit and of the microdosimetry subsystem in LURAD.

## 3 SIMULATIONS

The response of the LURAD instrument to the radiation of the lunar environment has been studied using Monte Carlo simulations, which have been performed with the GEANT4 toolkit version 4.10.06.p02 (Agostinelli, 2003). The QGSP_BERT_HP physics list has been used, which allows to simulate reactions of high energy hadrons with nuclei and to simulate high energy electro- and gamma-nuclear reactions. In addition, the "Bertini cascade model" treats nuclear reactions initiated by long-lived hadrons and gammas with energies between 0 and 10 GeV, while "High precision neutron model" uses cross-sections for neutrons with energies up to 20 MeV from the ENDF/B-VI evaluated data library.

The response of the proton-ion-fast neutron and electron-gamma subsystems has been studied independently from each other. Each subsystem was placed at the center of a sphere with radius 20 cm, which was emitting particles to its interior with equal probability from all the points of its internal surface. The direction of the emitted particle was selected randomly with probability proportional to the cosine of the angle between the direction and the radius at that point. The number of events generated was $2\cdot10^6$ for protons, alphas, neutrons, gammas, and electrons and varying from $1\cdot10^5$ to $2\cdot10^6$ for ions with Z>2. The energy distributions used were: (a) For the proton events the combined GCR and lunar albedo. (b) For electron, positron, gamma photon and neutron events the albedo energy distributions. (c) For ion with Z ≥2 events their GCR energy spectra. In the mixed field, the contribution of each particle type was scaled based on the integral of the particle's flux over energy. With this scaling the appropriate weight was assigned to each particle type in the weighted sum for all particles. The energy depositions into each active volume of the two subsystems have been used to study the performance of the system regarding its particle identification capability and energy estimation.

The simulated response of the SOI microdosimetry sensors in various space radiation environments has already been reported. Comparisons with experimental measurements performed in beam tests have shown good agreement with simulations. The simulated performance in the Columbus module of the ISS has been reported in (Peracchi et al., 2019). Comparisons between experimental and simulated results for high energy ions typical for GCR have been reported in (Peracchi et al., 2020, Tran et al., 2018) for a proton beam mimicking SPE in (Peracchi et al. 2021) and in fast neutron field in (Pan V.A. et al., 2022, Vohradsky et al., 2020). These results provide enough confidence for the performance of the microdosimetry sensors in the lunar field.

## 3.1 Radiation Field on the Lunar Surface

The GCR model used is the semiempirical International Organization for Standardization (ISO) 15390 model (ISO, 2004) which is based on models developed at Moscow State University (Nymmik and Suslov, 1995; Nymmik et al., 1996). With the aid of SPENVIS, GCR differential flux energy spectra of the 1996.4 solar minimum at one astronomical unit (AU) without geomagnetic shielding was produced. The GCR flux was divided by 2 to account for the shielding of half the sky (2π sr) by the Moon. Based on data provided by the Luna-16 and Luna-20 vehicles (Denisov, 2011), the lunar regolith has been divided into 5 layers. The depths of all layers and their density are summarized in Table *3*. All layers have identical chemical composition which is presented in Table *4*.

Table 3: Depths and densities of the Lunar Regolith model used in the Geant4 simulation

| Layer | Depth (cm) | Density (g/cm$^3$) |
|---|---|---|
| 1 | 0-0.5 | 0.6 |
| 2 | 0.5-20 | 1.2 |
| 3 | 20-35 | 1.5 |
| 4 | 35-500 | 2.0 |
| 5 | 500-1000 | 3.4 |

Table 4: Chemical Composition of the lunar regolith

| Element | Mass Percentage (%) |
|---|---|
| $^8$O | 42.53 |
| $^{12}$Mg | 4.84 |
| $^{13}$Al | 6.9 |
| $^{14}$Si | 19.7 |
| ²⁰Ca | 8.1 |
| ²²Ti | 4.33 |
| ²⁶Fe | 13.6 |

A hemispherical radiation source with 20 m radius (see Figure 7) has been placed on top of a stack of cylindrical sectors with the depths, mass densities given in Table 3 and chemical composition given in Table 4. The particles are emitted with a random direction in the interior of the sphere with a probability that is proportional to the cosine of the angle between the direction and the radius at that point. $10^5$ events were generated for each ion species with atomic numbers from Z=1 to Z=26. The particles emanating from the lunar surface were counted. The energy distribution of an albedo particle results as the sum of the scaled energy distributions of this particle arising from the Monte Carlo experiments for each one GCR ion. The scaling factor is the relative contribution of each specific ion in GCR.

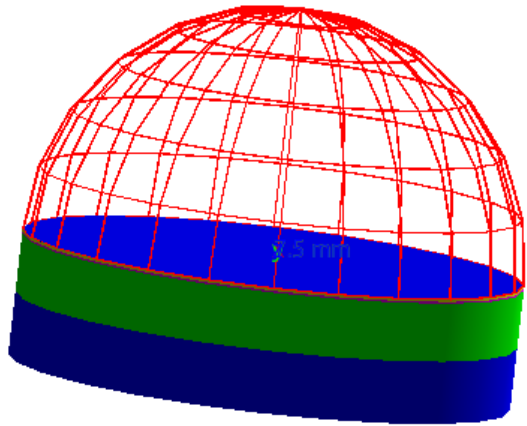

Figure 7: The Lunar model used in the Gean4 simulation for the estimation of the albedo particles: A hemispherical radiation source on top of a stack of 5 cylindrical regolith layers with radius of 20 m.

The particles emanating from the top surface of the cylindrical lunar regolith model are counted and their flux is calculated with the aid of the equation below:

$$albedo\_flux = \frac{(number\ of\ albedo\ particles\ from\ 10^5\ GCR\ evts)}{10^5} \times \frac{1}{2}(total\ GCR\ fux\ integrated\ over\ 4\pi\ sr)$$

The resulting relative contribution of the particle species is given in Table 5. Differential and integral energy spectra are presented in Figure 8.

Table 5: Relative contribution of Albedo particles from GEANT4 simulation

| Particle | % Contribution |
|---|---|
| gamma | 65,66 |
| neutron | 30,64 |
| e- | 1,294 |
| Proton (backscattered) | 0,925 |
| e+ | 0,609 |
| deuteron | 0,028 |
| mu+ | 0,0113 |
| triton | 0,0057 |
| mu- | 0,0050 |
| Other particles with very short lifetime. Amongst them pi+ and pi- which may increase slightly the flux of muons and electrons | 0,8 |

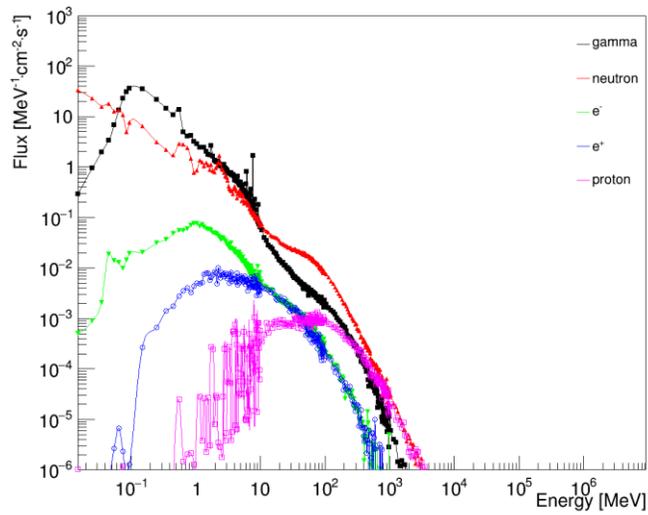

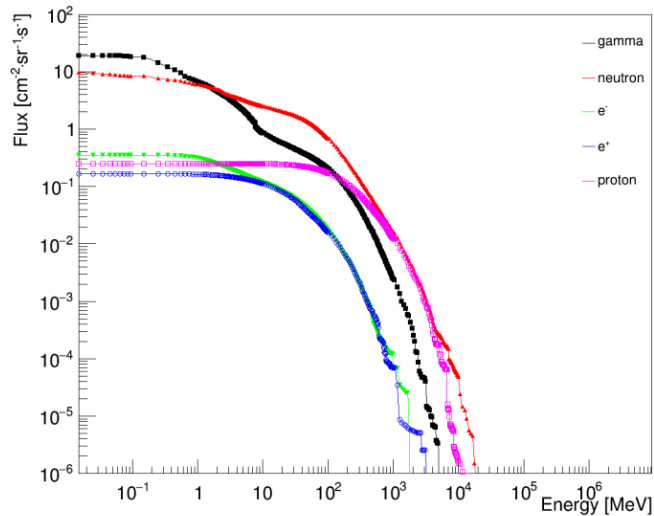

Figure 8: Differential (top) and integral (bottom) flux spectra of the abundant albedo particles on the lunar surface.

## 3.2 MULTIVARIATE ANALYSIS

The Toolkit for Multivariate Data Analysis with ROOT (Hoecker A. et al., 2007) has been used to apply multivariate classification and regression to the simulated response data of the proton-ion-fast neutron and the electron-gamma subsystems. Multi-Layer Perceptron Artificial Neural Networks (MLP) have shown the best performance for the identification of neutrons against all other particles present in the lunar radiation field. Gradient Boosted Decision Trees (GBDT) proved quite successful in classifying particles in the GCR field according to their atomic number and in the determination of the proton kinetic energy. For completeness, the MVA methods used are briefly presented.

A Multi-Layer Perceptron Artificial Neural Network (MLP) (Bishop C, 1995) consists of interconnected nodes, called neurons, which are organized in layers. Signals travel from the first layer (input) to the last layer (output). The internal layers are known as hidden layers. Each neuron provides to the following nodes the output of its activation function whose argument is the dot product of the vectors of inputs and of the weights plus the bias. The weights and the biases are subject to adjustment achieved through the training. After training, a mapping from a space of input variables onto a space of output variables has been constructed. By applying an external signal, the network is put into a defined state that can be measured from the response of one or several output neurons.

A Gradient Boosted Decision Tree (GBDT) (Hastie T et al*., 2009)* is a binary tree structured classifier or regressor. Repeated yes/no decisions are taken on one single variable at a time until a stop criterion is fulfilled. The phase space is split this way into many regions that are eventually classified as signal or background, depending on the majority of training events that end up in the final leaf node. With boosting a forest of trees is built. The additional trees are derived from the same training ensemble by reweighting events and are finally combined into a single classifier or regressor which is given by a weighted average of the individual trees. Gradient boosting means that the addition of trees for the minimization of the error is done with a gradient descent procedure rather than giving a higher weight to misclassified events.

## 4 SIMULATED PERFORMANCE OF LURAD SUB-SYSTEMS

The design of the LURAD sub-systems has been optimized with the aid of iterations of simulation experiments in which the response in the realistic lunar field was recorded and subsequently analyzed with purpose to understand the factors affecting the efficiency and purity of particle identification and the fidelity of the kinetic energy spectra determination.

### 4.1 PROTON-ION-FAST NEUTRON SUBSYSTEM

The proton-ion-fast neutron sub-system should be capable of identifying charged particles and neutrons while on the other hand reject gammas and electrons. The lunar albedo field contains gamma photons with a flux around two times higher than the fast neutron flux (see Table *5*). As a result, the detector should have a high gamma background rejection efficiency when performing the neutron identification.

#### 4.1.1 Neutron identification and kinetic energy reconstruction

Neutral particle candidate events are those which simultaneously: (a) produce signal on one or more scintillator volumes of the central sensitive cube (D and E in Figure 2); (b) do not produce signal on the two low gain Si detectors placed above and below the central sensitive cube (B1 in Figure 2) and (c) do not produce signal on the surrounding EJ200 plastic anticoincidence detectors (F in Figure 2). In the case of neutral particles (neutrons, gammas) the detectors B1 and F act as anticoincidence detectors. Only neutral particles can go across them without depositing energy, while the charged particles are rejected. The sample that satisfies conditions (a) through (c) is dominated by gammas as it is expected due to their abundance in the lunar albedo field. Specifically, it contains 89% gamma photons, 10% neutrons and 1% other particles. Neutrons interact mainly with the central detector (detector D) and due to elastic scattering, transfer part of their energy to the protons which are the nuclei of the hydrogen atoms of the plastic detector D. If criterion (a) is modified to include events that necessarily involve energy deposition on the EJ200 scintillator labeled D in Figure 2, then the neutron component becomes dominant. The recoil protons may escape from detector D and produce signal on the surrounding CsI detectors (detectors E). With purpose to collect the energy of all the produced protons, the total energy deposited on D and E is taken into account. More than 75% of the events fulfilling the modified criterion (a) and criteria (b) and (c) are neutrons. Many gamma photons interact with the high-Z CsI detector E, but their secondaries do not reach detector D. The absolute efficiency defined as the neutrons correctly identified over the total number of neutrons impinging to the detectors is 1,26%.

The capability for the reconstruction of the neutron kinetic energy was tested using a sample of neutron events having the albedo energy distribution. For each event, the energy depositions in scintillators D and E were added and the resulting distribution was input to the method described in (Lambropoulos C.P. et al., 2019). The reconstructed and the original kinetic energy distributions are compared in Figure 9.

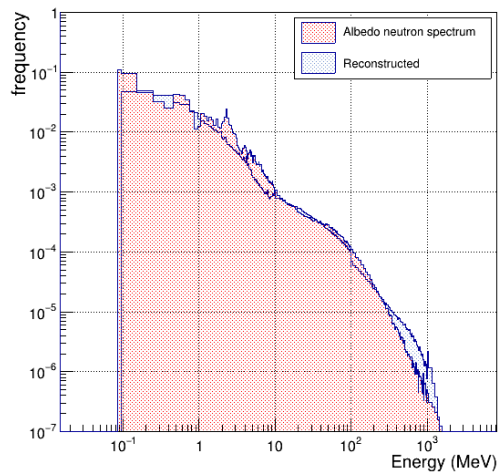

Figure 9: Reconstructed spectrum of neutron kinetic energy compared to the albedo neutron energy distribution.

In addition, neutrons were identified with the aid of an MLP classifier having as input variables the energy depositions on all the detector volumes except the low gain Si and the plastic anticoincidence detectors. The MLP was trained using a sample with the particle composition and energy distribution of the lunar radiation field, encompassing both GCR and albedo particles. The classifier efficiency was evaluated on an independent sample. By requesting the neutron neural net output variable to be greater than 0.9, 79% of the events were correctly classified as due to neutrons, while the rest were gamma particles incorrectly identified as neutrons.

### 4.1.2 Proton and alpha particles identification

As charged particle candidate events are considered those that deposit energy on both the low gain Si detectors (B1 in Figure *2*). However, this requirement excludes low energy particles stopping in the central scintillator cube. This case is discussed in section 4.1.5. for protons. If protons and alpha particles are sought, then events producing signal on the surrounding EJ200 plastic anticoincidence detectors are rejected. Events giving signal on both the B1 Si pixel layers are mainly protons (89% of the sample) and secondarily alpha particles.

Identification was performed using GBDT. Input variables were the deposited energies on the Si Pixel layers and on the scintillators D and E. The algorithm produces four outputs, one for each particle class that identifies i.e., proton, alpha, electron, and heavier ions. Its performance was evaluated on an independent sample. By requesting the proton output variable to be greater than 0.7, 98% purity was obtained, the main contamination being from electrons misidentified as protons. By requesting the alpha particle output variable to be greater than 0.7, 95% purity was obtained. In this case the main contamination was from protons. The resulting probability of correct identification for protons is 0.97 while for the alpha particles is 0.94.

### 4.1.3 Proton kinetic energy estimation

Kinetic energy estimation limits for protons were produced using relativistic analytical relations. These limits present the estimated kinetic energy versus the real one when the measured time-of-flight is in the $[\frac{L}{v} + \sigma, \frac{L}{v} - \sigma]$ interval, where L the travel distance, v the proton speed and σ the standard deviation of the time difference measurement. The results indicate that the estimated energy for each event may deviate significantly from the real one with increasing real energy and decreasing time resolution.

Using the simulated detector response, the proton kinetic energy estimation was handled as a regression problem to be solved with the aid of GBDT. The GBDT input variables were: (a) the energy depositions on scintillators D and E; (b) the energy depositions on all the Si Pixel detectors; (c) the time difference between the signals of the same track on the two fast scintillators. The time difference was smeared to consider the effect of finite measurement resolution using a Gaussian distribution with sigma of 10 ps, 30 ps or 50 ps.

In Figure 10(a) is presented the profile plot of the real kinetic energy as a function of the estimated one for 30 ps smearing of the time difference. In the profile plot the MC events are collected in bins of estimated energy with a width of 50 MeV each. The points represent the mean value of the real energy and the bars the uncertainty in the estimation of the mean. The estimated kinetic energy spectrum for events selected as protons and with 30 ps smearing of the time difference is compared to their Monte Carlo truth spectrum in Figure 10(b). The two spectra are quite similar. This result implies that the probability of underestimating proton kinetic energy is almost the same with the probability that it is overestimated, although the energy extracted for each individual proton event has higher uncertainty with increasing energy value.

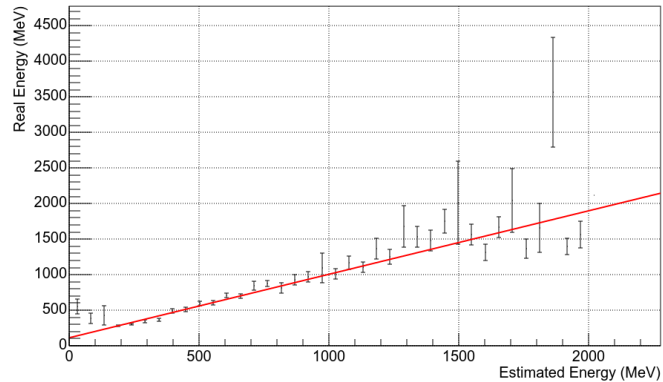

(a)

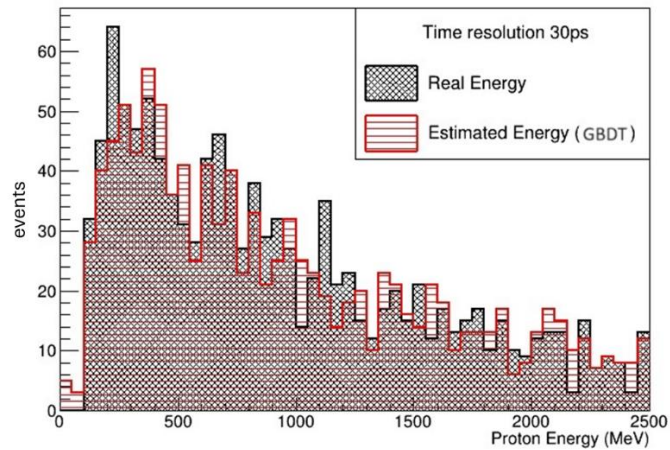

(b)

Figure 10: (a) Profile plot of the proton real energy as a function of the estimated one for 30 ps time measurement resolution. (b) Estimated proton energy spectra compared to the real one for 30 ps time measurement resolution.

### 4.1.4  Identification of ions with Z>2

To clearly identify heavier ions, it's necessary to measure the energy deposited on the B1 Silicon (Si) pixel layers. It's important not to dismiss events that generate signals on the anticoincidence scintillators, despite any initial inclination to do so. This is because ions, when within the detector's field of view, produce delta rays that directly impact the anticoincidence scintillators. Consequently, rejecting such events would inadvertently remove the majority of the heavy ion tracks, which are crucial for study and observation. Moreover, the energy deposited on the anticoincidence detectors can be harnessed to enhance the capability to identify ions more accurately. Furthermore, while the arrangement of the energy depositions on the Si pixel detectors can be used to identify heavy ions, it is worth noting that this potential strategy has not yet been thoroughly explored or studied.

The atomic number estimation of the charged particle events not identified as protons or alphas is handled as a regression problem with GBDT. Energy depositions of all volumes (including the EJ200 anticoincidence detectors) have been used as inputs to the GBDT with output the ion atomic number. Training was performed with particles having the energy distribution of the lunar field (GCR + lunar albedo), but their number did not correspond to their relative contribution in the lunar field, because for heavier ions the sample size would be extremely low. An independent set of simulated data has been used to evaluate the atomic number estimation of the GBDT algorithm. The atomic number, Z, arising as the output of the GBDT algorithm has been rounded to the nearest integer value. This value is considered as the estimated atomic number Z.

To estimate the probability of accurately identifying ions, the number of events for each ion sample was scaled to half of the ion's integral flux in the Galactic Cosmic Ray (GCR) spectrum. After applying the ion identification algorithm, the number of correctly identified ions is compared to the total number of ions identified to belong to that species. The probability of identifying a specific ion is defined by the ratio of these two numbers. Figure *11* illustrates the probability of correctly identifying an ion in relation to its actual atomic number (Z), considering ions with the GCR energy distribution. The notably low probability of accurately identifying Fluorine (F), Phosphorus (P), Scandium (Sc), and Manganese (Mn) - with atomic numbers 9, 15, 21, and 25 respectively - arises from their substantially lower flux compared to neighboring ions. Misidentifying the high-flux neighboring ions as F, P, Sc, or Mn significantly contaminates the F, P, Sc, and Mn sample, leading to a diminished probability of accurately identifying these ions.

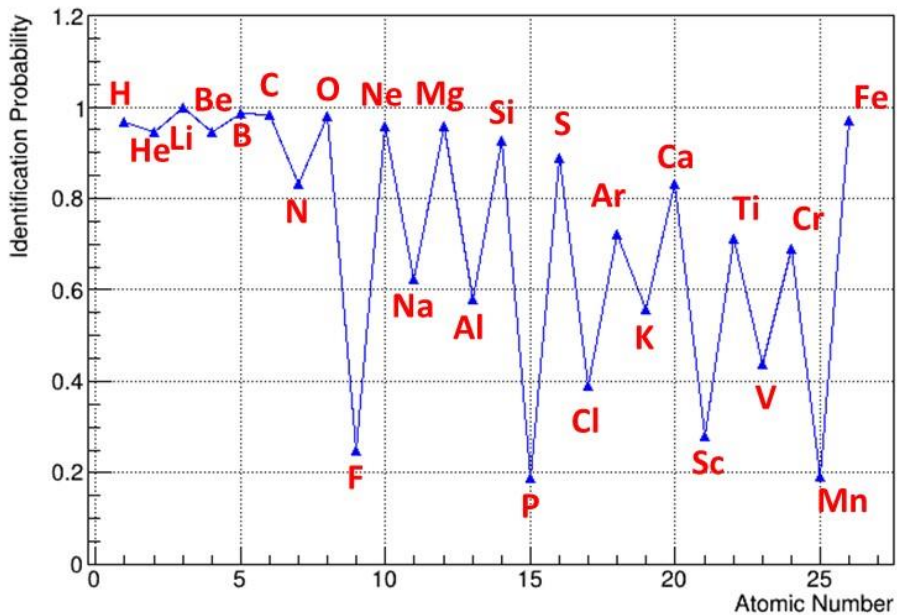

Figure 11: Probability of correct identification of ions with atomic number Z=1 to Z=26 as a function of the atomic number. The plot included the probability of correct identification of proton and alpha particles with the algorithm described in paragraph 4.1.2 and the probability of correct identification of heavier ions with Z>2 as described in paragraph 4.1.4

### 4.1.5 Low energy protons

Low energy protons impinging on the detector should stop inside the scintillator cube and consequently should deposit energy on only one of the low gain Si detectors. It was noticed that, indeed, this condition is satisfied by the proton events with primary energy below 150 MeV, but also a significant number of events due to protons with energy higher than 150 MeV contributes to the sample (see Figure *12*(a)). It was verified that this happens because high energy protons impinging on the lateral surfaces of the detector interact with non-active volumes and produce secondary electrons which satisfy the selection criterion. However, these secondary particles deposit relatively low energy on the scintillator cube, as one can see in Figure *12*(b). The energy deposition spectra presented in Figure *12*(b) suggest that with a cut of 40 MeV most of the proton events with energy higher than 150 MeV are rejected. However, with this cut proton events with energy below 40 MeV were also rejected. For the estimation of the kinetic energy of protons with energies below 40MeV a better algorithm remains to be developed.

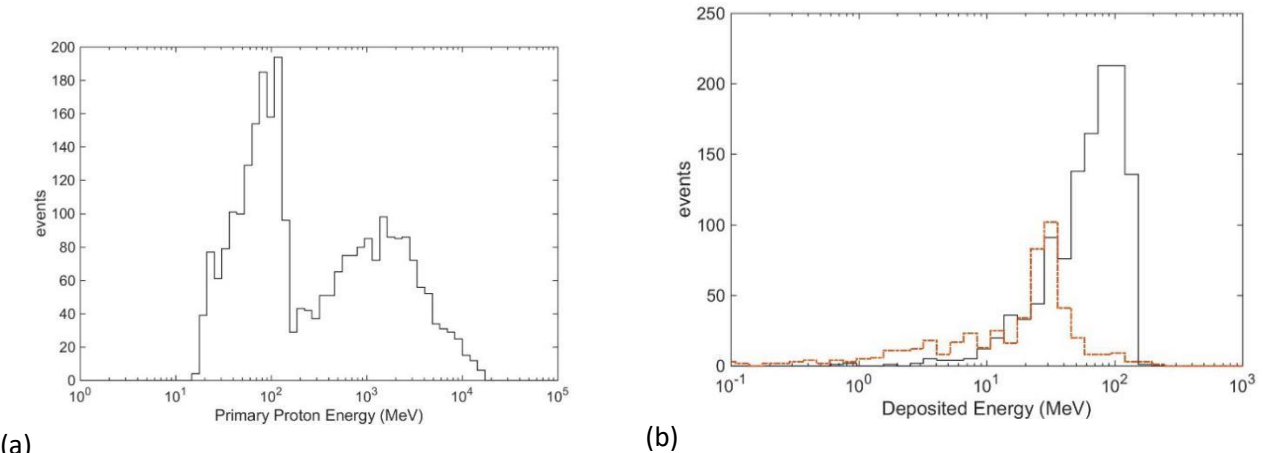
(a) (b)

Figure 12: (a) Kinetic energy spectrum of protons that deposit energy on only one low gain Si sensor and on the central scintillator cube. (b) Spectrum of energy deposition on the central scintillator cube for protons with kinetic energy in the region below 150 MeV (black line) and above (red dashed line).

## 4.2 Electron-gamma subsystem

A peculiarity of the lunar albedo field is that the gamma photons flux is two orders of magnitude higher than the electron / positron component. This abundance of gamma photons is predicted by the results of Monte Carlo simulations of the interaction of galactic cosmic rays with the lunar regolith, as can be seen in Figure *8*. Protons impinging on the lunar surface produce gamma photons and electrons. Photons can escape from deeper layers of the regolith than electrons. So, a much higher photon flux than the electron one results on the lunar surface. Consequently, the detector should have a very high gamma background rejection efficiency when performing the electron identification.

### 4.2.1 Electron selection

The main criterion for the electron selection is to count the events where signal is produced on the three Si pixel detectors and possibly on the central CsI(Na) cube ( ScID=1 in Figure *3*) and reject all the events where energy is deposited on the anticoincidence CsI(Na) volumes ( ScID=2, 3,4,5,6 in Figure *3*). This implies that when the instrument operates on the lunar surface, the PixID=1 Si layer should face the lunar regolith.

As the study revealed, a significant contamination of the electron signal arises from photons undergoing pair production near the surface of the ScID=1 crystal. Such photon events produce electrons and positrons which deposit energy on the Si pixel detectors. It is possible to reject most of these events because electrons and positrons produce spatially separated clusters of energy depositions on the pixel detectors. Consequently, the criterion imposed was that events should have one energy cluster on each of the Si pixel detector layers next to the ScID=1 crystal (PixID=3 and PixID=2 in Figure *3*).

Protons that pass through the Si pixel layers and stop in the ScID=1 crystal could contaminate the electron signal, but it has been understood that by limiting the energy deposition on ScID=1 to be less than 18 MeV reduces this contamination. Furthermore, if energy deposition less than 150 keV on the Si pixel layer with PixID= 1 is required, then the result is the rejection of most protons.

The number of electron-induced events was estimated to be ≈ 83% of the total events that pass all the imposed criteria. The electron-events that survive the additional criteria is almost the 66% of those electron-events that fulfil the main criterion described above.

As electron kinetic energy has been taken the sum of energy depositions on the three Si pixel layers and on the ScID=1 crystal. In Figure *13* is compared the resulting spectrum with the true kinetic energy of the primary particles of the events used.

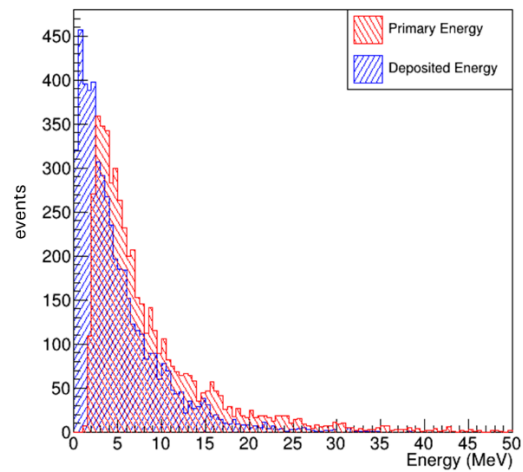

Figure 13: The kinetic energy spectrum of the primary particles identified as electrons in red and the calculated spectrum from the energy depositions in blue.

### 4.2.2 Gamma selection

Events due to gamma photons are considered those that deposit energy only on the ScID=1 crystal and do not produce signal on the three Si pixel detectors. The sample selected using this criterion is significantly contaminated by neutron events, but it is observed that neutrons deposit much lower energy on the the ScID=1 crystal than the gamma photons. Thus, if only events with energy deposition ≥50 keV are accepted, the purity of the gamma events sample reaches 93%. Almost 97% of gammas survive the 50 keV cut.

For the events classified as gamma rays, the energy deposition on the ScID=1 crystal is considered as the energy of the gamma ray. In Figure *14*, the energy deposition spectrum on the ScID=1 crystal of these events is compared with the true energy spectrum of the gamma rays in the sample.

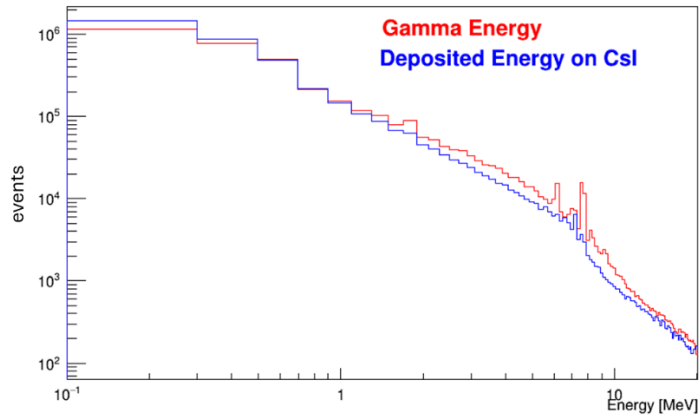

Figure 14: In red, energy spectrum of the true gamma photons in the sample. In blue, the spectrum of measured energy deposition in the CsI(Na) for the same sample.

## 5  CONCLUSIONS - DISCUSSION

The primary purpose of the reported work is to present the design and explore the measuring capacity of an instrument with low mass footprint performing in the lunar environment.

In the presented study of LURAD capabilities, some factors which could not be evaluated quantitatively at the current stage of development were not considered. These are mainly the following:

If the particle passing through the detector is a proton, then energy deposition most probably occurs along its path. But if the particle is a heavy ion, then delta rays (electrons) are emitted with a considerable path length and the energy deposition occurs also at distant positions from the ion's path. Each energy deposition results to electron-hole creation and transfer to the electrodes with drift due to the electric field and with diffusion. Consequently, the conversion of energy depositions in the Si active pixel sensors to charge signal induced on the pixel electrode depends on the track within the Si bulk, on the species of the charged particle and on the electric field distribution. The response of the pixel electronics will be non-uniform to a certain extent and affected by noise. The inclusion of such effects in the simulations requires extensive characterization of the active pixel sensors.

The time-of-flight system should have the capability to measure time differences between 1 to 2 ns with a resolution of 30 ps or better. These specifications have been set after estimating the uncertainty in kinetic energy determination as a function of the time resolution. For simplicity, in our Monte Carlo study, we have assumed a constant time resolution of 50 ps, 30 ps, and 10 ps and we have concluded that 30 ps is a good compromise considering the current state of the art (Stoykov A. and Rostomyan T., 2021). Resolution is governed by a relation of the form $\sigma = \sigma_{1MeV}/\sqrt{E}$ where E is the recorded energy deposition on the scintillator and $\sigma_{1MeV}$ should be optimized.

It is well known, that when a proton or an ion passes through the plastic scintillator, the number of optical photons produced has a non-linear relation with the energy deposited. This phenomenon, parametrized as Birks law, has the result that the same deposited energy from protons and other ions produces lower light output compared to the one which arises from electron passage.

Nevertheless, even with the mentioned limitations of the study, the simulated response of the proton-ion-fast neutron and electron-gamma subsystems indicates that LURAD can perform identification and spectroscopic measurements for all the main energetic particles present in the mixed radiation field of the lunar environment.

Proton, alpha and heavier ion atomic numbers can be discriminated using the GBDT method. The study has been performed for ions with Z up to 26. It has not been extended to ions with higher Z, because the probability for the instrument to register ions with Z>27 is very low. The kinetic energy estimation has been studied for protons and the results show that proton energy spectra can be reproduced with fidelity. Fast neutrons can be identified effectively. The purity of the sample of events identified as neutrons is greater than 75% and the absolute efficiency of detection is 1.26%. Using the spectrum unfolding method described in (Lambropoulos et al., 2019), the reconstructed neutron kinetic energy spectrum is in accordance with the original one at least for energies from 100 keV up to 200 MeV. The particle identification capability and the extension of kinetic energy determination above 500 MeV place LURAD beyond the state of the art, at least at the design level, with a low mass footprint.

LURAD performs well for the measurement of the photon flux as a function of energy in the range from 100 keV up to 20 MeV. Photons can be identified with purity 93%. The reconstructed gamma energy spectrum matches well with the initial one. Electrons can be effectively identified. Even though the electron flux is a tiny component of the lunar field (less than 1%) their abundance in the events sample isolated using the imposed criteria is 83%. However, the spectrum of deposited energy on the volumes used for the energy determination is shifted towards lower energies compared to the original. The sensitivity in electron identification may render LURAD valuable in heliophysics investigations at the Gateway orbit (see Ameri D. et al. 2019, Mewaldt R.A. et al., 2005).

## ACKNOWLEDGMENTS

This work has been funded by the European Space Agency Contract 4000133574/21/NL/CRS for the development of a Comprehensive radiation monitor package for Lunar mission. Also, it has been supported by computational time granted from the Greek National Infrastructures for Research and Technology S.A. (GRNET) in the National HPC facility – ARIS.